\documentclass[conference]{IEEEtran}

\usepackage{amsmath,amssymb,bm}
\usepackage{cite}
\usepackage{graphicx}
\usepackage{tikz}
\usepackage{array}
\usepackage{booktabs}
\usepackage{url}
\usepackage{stfloats}

\newcommand{\CN}{\mathcal{CN}}

\newcommand{\C}{\mathbb{C}}
\newcommand{\bw}{\bm{w}}
\newcommand{\blam}{\boldsymbol{\lambda}}
\newcommand{\Ptot}{P_{\mathrm{total}}}
\newcommand{\lmin}{\lambda_{\min}}
\newcommand{\Nmc}{N_{\mathrm{mc}}}

\title{LLM-Steered Power Allocation \\for Parallel QPSK-AWGN Channels}

\author{
\IEEEauthorblockN{Tadashi Wadayama}
\IEEEauthorblockA{Nagoya Institute of Technology}
}
\begin{document}
\maketitle

\begin{abstract}
  Large language models (LLMs) are increasingly being explored as high-level decision modules 
  in closed-loop systems, but their stochastic nature makes safe integration challenging. 
  In this paper, we propose \emph{LLM-Steered Power Allocation}, a dual-process architecture for parallel QPSK 
  channels inspired by Kahneman's System~1/System~2 framework. A fast numerical optimizer (System~1) 
  continuously performs projected gradient ascent on a weighted mutual-information objective, 
  while an LLM navigator (System~2) periodically interprets natural-language policies and updates 
  only the channel weights and the operational power budget. The LLM never manipulates 
  the power-allocation variables directly, and constraint satisfaction is enforced structurally 
  by the optimizer. To mitigate LLM unreliability, we further incorporate multi-layer guardrails 
  including normalization, exponential moving-average smoothing, and fallback mechanisms. 
  Numerical experiments on an 8-channel system show that, with a fixed optimization core and unchanged system 
  prompt, different natural-language policies induce qualitatively different operating points, 
  including throughput-oriented allocation, channel prioritization, power-aware operation, 
  and channel shutdown. In addition, under an abrupt channel-gain reversal, the proposed system 
  autonomously reconfigures its steering signals and reduces the final mutual-information 
  spread by 60\% compared with the optimizer alone. These results suggest that LLMs 
  can serve as policy interpreters for safe, flexible reconfiguration of communication-system optimizers 
  without controller reimplementation.
\end{abstract}


\section{Introduction}\label{sec:intro}

Large language models (LLMs) are increasingly explored as decision
modules in wireless communications---for network management, protocol
design, and resource
optimization~\cite{llm_agent_survey, react, llm_telecom_survey,
llm_networking, llm_wireless}.
Most existing work, however, employs LLMs for one-shot or offline
tasks.
Recent work has also applied LLMs directly to resource allocation
in wireless systems~\cite{noh_llm_ra}, but without an explicit
mechanism to prevent constraint violations caused by erratic LLM
outputs.
The key open question is therefore not whether an LLM \emph{can}
suggest a control action, but how to integrate one into a
\emph{real-time, closed-loop} optimizer without letting its
inherently stochastic outputs violate system constraints.

A key goal in next-generation wireless systems is
\emph{resilience}~\cite{resilient_wireless, shen_edge_llm}: the ability to maintain or
swiftly recover performance under unforeseen environmental changes---such
as abrupt channel fading, traffic surges, or hardware
degradation---without requiring human re-intervention.
Conventional optimizers can track a moving optimum numerically, but they
cannot \emph{reinterpret} the situation: when channel conditions change
qualitatively, the optimization \emph{objective itself} may need
revision (e.g., switching from throughput maximization to fairness-aware
allocation).
This kind of semantic, context-dependent re-evaluation is one
motivation for using LLMs as high-level policy interpreters.
By leveraging an LLM's ability to reason about system state in the
context of high-level operator intent, the system can adapt not only
\emph{how} it optimizes but \emph{what} it optimizes for.
This motivates placing an LLM \emph{inside} the control loop as a
high-level navigator that steers the optimization strategy accordingly.

Power allocation over parallel additive white Gaussian noise (AWGN)
channels is a fundamental problem in wireless communications and serves
as a natural testbed for LLM-in-the-loop control.
The parallel-channel model directly underlies
practical multi-band systems such as OFDM and carrier aggregation,
so insights obtained here carry over to real-world deployments.
When the input constellation is discrete---such as QPSK ($M=4$)---the
mutual information (MI) per channel saturates at $\log_2M = 2$~bits,
making the classical water-filling solution suboptimal and rendering
closed-form solutions unavailable~\cite{lozano_mercury}.
Numerical optimization is therefore required, and the optimal allocation
depends on channel conditions that may vary over time.
This combination of numerical difficulty, time-varying optimality, and
the need for flexible, multi-objective trade-offs (e.g., throughput vs.\
fairness) makes QPSK parallel channels a compelling---yet deliberately
simple---setting in which to isolate and evaluate an LLM steering
mechanism.

In this paper, we propose \emph{LLM-Steered Power Allocation} for
this problem.
Our architecture is inspired by Kahneman's {dual-process
theory}~\cite{kahneman}: System~1 denotes fast, automatic cognitive
processes, while System~2 denotes slow, deliberative reasoning.
We map this dichotomy onto the power allocation system:
a fast projected-gradient optimizer ({System~1}) continuously
maximizes a weighted sum MI objective under power constraints, while
an LLM ({System~2}) periodically evaluates the system state against
a user-specified natural-language \emph{policy} and adjusts the
optimization objective and power budget accordingly.
The two systems operate \emph{asynchronously} at different time scales,
so that LLM latency never blocks the optimizer.

Unlike rule-based controllers (which need per-policy code changes) or
reinforcement learning (which needs retraining), an LLM interprets a
new natural-language policy at run time without any
reimplementation---enabling zero-shot adaptation to diverse intents.

Our contributions are threefold:
\begin{enumerate}
\item A \emph{dual-process architecture} (System~1\,/\,System~2)
  for LLM-steered power allocation with bounded, indirect actuation
  that ensures feasibility by construction---the LLM cannot violate
  power constraints by design.
\item A set of practical \emph{guardrails}
  (normalization, EMA smoothing, fallback, and explainability)
  that mitigate malformed or erratic LLM outputs and enable
  graceful degradation under LLM failure.
\item \emph{Experimental evidence} that natural-language policies
  steer the optimizer to distinct operating points and support
  policy-driven adaptation under channel variation.
\end{enumerate}

An interactive demo is available online.\footnote{\url{https://github.com/wadayama/llm-steered-power-allocation}}
The remainder of the paper is organized as follows.
Section~\ref{sec:prelim} introduces the system model and mutual
information estimation.
Section~\ref{sec:method} presents the proposed dual-process architecture
and its guardrails.
Section~\ref{sec:experiments} provides numerical results, and
Section~\ref{sec:conclusion} concludes with future directions.

\section{Preliminaries}\label{sec:prelim}

\subsection{System Model}\label{sec:system_model}

We consider $N$ parallel, independent sub-channels, each carrying
QPSK-modulated symbols.
We focus on QPSK ($M=4$) to keep the presentation concise;
the framework extends straightforwardly to larger constellations
such as 16-QAM or 64-QAM by replacing the constellation set~$\mathcal{S}$.
The input--output relationship on sub-channel~$i$ ($i=1,\dots,N$) is
\begin{equation}\label{eq:channel}
  Y_i = h_i \,\lambda_i \,X_i + Z_i,
\end{equation}
where $h_i \in \C$ is the known complex channel gain,
$\lambda_i \ge 0$ is the transmit amplitude scaling factor
(the allocated power is $P_i = \lambda_i^2$),
$X_i$ is drawn uniformly from the QPSK constellation
$\mathcal{S} = \{(\pm 1 \pm j)/\sqrt{2}\}$ ($M=4$ points),
and $Z_i \sim \CN(0,\sigma^2)$ is additive white Gaussian noise.
Since $h_i$ can take arbitrary complex values, we set $\sigma^2 = 1$
without loss of generality.
The total power is subject to the constraint
\begin{equation}\label{eq:power_constraint}
  \sum_{i=1}^{N} \lambda_i^2 \le \Ptot, \qquad \lambda_i \ge 0,
  \;\; i=1,\dots,N.
\end{equation}
Here $\Ptot$ is treated not as a fixed hardware limit but as an
\emph{operational power cap} that may be adjusted at run time to
reflect battery constraints, thermal limits, or interference
management policies.
This interpretation motivates the design choice in
Section~\ref{sec:system2}, where the LLM is allowed to modify $\Ptot$
as part of its control action.

\subsection{Mutual Information for QPSK Input}\label{sec:mi_estimation}

To express the MI in a computable form, we work with real-valued
representations.
Since $Z_i \sim \CN(0,\sigma^2)$, each real and imaginary component has
variance $t \triangleq \sigma^2/2 = 1/2$.
Let $a_i = |h_i|\,\lambda_i$ denote the effective amplitude on
sub-channel~$i$,
$\tilde{s}_k = [\operatorname{Re}(s_k),\,\operatorname{Im}(s_k)]^\top$
the real representation of constellation point $s_k \in \mathcal{S}$,
$\Delta_{kj} = \tilde{s}_k - \tilde{s}_j$, and
$\tilde{Z} \sim \mathcal{N}(\bm{0},\, t\,\bm{I}_2)$ the real noise vector.
Throughout this paper, $\log$ denotes the base-2 logarithm and all
information quantities are measured in bits.
The MI of sub-channel~$i$ under QPSK input is given by~\cite{cover_thomas}
\begin{align}\label{eq:mi_exact}
  I(X_i;Y_i) &= \log M \notag\\
  &- \frac{1}{M}\sum_{k=1}^{M}
  \mathbb{E}_{\tilde{Z}}\!\left[
    \log \sum_{j=1}^{M} e^{\,\psi_{kj}}
  \right]\!,
\end{align}
where
\begin{equation}\label{eq:psi}
  \psi_{kj} \triangleq
    -\frac{\|a_i \Delta_{kj}\|^2}{2t}
    -\frac{a_i \Delta_{kj}^\top \tilde{Z}}{t}\,.
\end{equation}
By drawing $Z^{(n)} \sim \mathcal{N}(\bm{0}, \bm{I}_2)$ and substituting
$\tilde{Z} = \sqrt{t}\,Z^{(n)}$, we obtain the Monte Carlo estimate
\begin{align}\label{eq:mi_mc}
  \hat{I}_i &= \log M \notag\\
  &- \frac{1}{M}\sum_{k=1}^{M}
  \frac{1}{\Nmc}\sum_{n=1}^{\Nmc}
  \log \sum_{j=1}^{M} e^{\,\hat{\psi}_{kj}^{(n)}}\,,
\end{align}
where
$\hat{\psi}_{kj}^{(n)} \triangleq
  -{{\|a_i \Delta_{kj}\|^2}}/{(2t)}
  -{a_i \Delta_{kj}^\top Z^{(n)}}/{\sqrt{t}}$.
The $\log\!\sum_j\!\exp(\cdot)$ is computed via the logsumexp trick for
numerical stability.
Since $a_i = |h_i|\,\lambda_i$ is a differentiable function of
$\lambda_i$, automatic differentiation (autograd) through the
computation graph of~\eqref{eq:mi_mc} directly yields
$\partial I_i / \partial \lambda_i$,
which we exploit for gradient-based optimization.
We write $I_i \triangleq I(X_i;Y_i)$ for brevity.
The maximum per-channel MI is $\log M = 2$~bits.

\subsection{Weighted Optimization Problem}\label{sec:weighted_problem}

A nonnegative weight vector
$\bw = [w_1,\dots,w_N]^\top$ with $\sum_{i} w_i = 1$
encodes the relative priority of each sub-channel.
The weighted sum MI objective is
\begin{equation}\label{eq:objective}
  \underset{\blam}{\text{maximize}}\;\;
  \mathcal{J}_{\bw}(\blam) \triangleq \sum_{i=1}^{N} w_i\, I_i
  \qquad
  \text{s.t.}\;\;\eqref{eq:power_constraint}.
\end{equation}
Since scaling all weights by a positive constant does not change the
optimizer $\blam^\star$, only the \emph{ratios} among the $w_i$
matter; we normalize to $\sum_i w_i = 1$.
Equal weights ($w_i = 1/N$) reduce~\eqref{eq:objective} to
sum-rate maximization.

\subsection{Why Water-Filling Is Suboptimal}\label{sec:waterfilling}

Under a Gaussian-input linear-channel approximation,
the weighted rate
$R_{\mathrm{lin}}(\bm{P}) = \sum_i w_i \log(1 + |h_i|^2 P_i / \sigma^2)$
is maximized by the weighted water-filling solution~\cite{cover_thomas}
\begin{equation}\label{eq:waterfilling}
  P_i^\star = \left[\frac{w_i}{\nu \ln 2}
              - \frac{\sigma^2}{|h_i|^2}\right]_+\!,
\end{equation}
where $\nu > 0$ is chosen so that $\sum_i P_i^\star = \Ptot$.

For QPSK input, however, MI saturates at $\log 4 = 2$~bits per channel.
In the high-SNR regime, water-filling over-allocates power to channels
that have already reached the saturation ceiling, wasting resources that
could improve weaker channels.
Consequently, water-filling is generally suboptimal for discrete-input
channels~\cite{lozano_mercury}, and no closed-form optimal solution is
available.
This gap motivates numerical optimization, and we use the water-filling
solution~\eqref{eq:waterfilling} as a familiar Gaussian-input
reference for comparison, rather than as a discrete-input benchmark.

\subsection{Gradient Ascent with Automatic Differentiation}
\label{sec:grad_ascent}

Since the MI estimate~\eqref{eq:mi_mc} is constructed entirely from
differentiable operations (scaling, exponentiation, logsumexp),
automatic differentiation (autograd) through its computation graph
directly yields the gradient
$\partial \mathcal{J}_{\bw}/\partial \lambda_i
 = w_i\,\partial I_i / \partial \lambda_i$
for each sub-channel.
This makes a projected gradient ascent approach straightforward to
implement: one simply evaluates~\eqref{eq:mi_mc} with a deep-learning
framework (e.g., PyTorch), calls \texttt{backward()}, and updates
$\blam$ in the gradient direction while projecting back onto the
feasible set~\eqref{eq:power_constraint}.

Note that $\mathcal{J}_{\bw}(\blam)$ is generally non-concave for
discrete inputs, so gradient ascent may reach a local rather than
global maximum; we have not observed this to be problematic for $M=4$.

\section{Proposed Method}\label{sec:method}

\subsection{Architecture Overview}\label{sec:architecture}

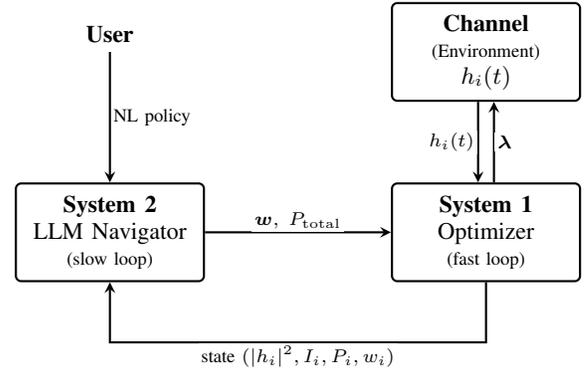
\begin{figure}[t]
\centering
\begin{tikzpicture}[
  block/.style={rectangle, draw, thick, rounded corners=2pt,
    minimum width=2.5cm, minimum height=1.3cm, align=center,
    font=\small},
  arr/.style={->, >=stealth, thick},
  lbl/.style={font=\scriptsize, fill=white, inner sep=1pt}
]
\node[block] (s2) at (0, 0)
  {\textbf{System 2}\\LLM Navigator\\{\scriptsize(slow loop)}};
\node[block] (s1) at (5, 0)
  {\textbf{System 1}\\Optimizer\\{\scriptsize(fast loop)}};
\node[block] (env) at (5, 2.4)
  {\textbf{Channel}\\{\scriptsize(Environment)}\\$h_i(t)$};
\draw[arr] (s2.east) --
  node[lbl, above] {$\bw,\;\Ptot$} (s1.west);
\draw[arr] (s1.south) -- ++(0,-0.8) --
  node[lbl, below] {\scriptsize state $(|h_i|^2, I_i, P_i, w_i)$}
  ++(-5,0) -- (s2.south);
\draw[arr] (0, 2.4) node[above, font=\small] {\textbf{User}} --
  node[lbl, right] {\scriptsize NL policy} (s2.north);
\draw[arr] ([xshift=-3pt]env.south) --
  node[lbl, left] {\scriptsize $h_i(t)$} ([xshift=-3pt]s1.north);
\draw[arr] ([xshift=3pt]s1.north) --
  node[lbl, right] {\scriptsize $\blam$} ([xshift=3pt]env.south);
\end{tikzpicture}
\caption{Block diagram of the proposed dual-process architecture.
The user provides a natural-language (NL) policy to System~2, which
steers System~1 indirectly through control parameters
$(\bw, \Ptot)$; it never touches the allocation $\blam$ directly.}
\label{fig:block_diagram}
\end{figure}

Our architecture, depicted in Fig.~\ref{fig:block_diagram}, consists of
two asynchronous loops inspired by Kahneman's dual-process
theory~\cite{kahneman}:

\begin{itemize}
\item \textbf{System~1 (fast loop, optimizer engine):}
  A projected gradient ascent algorithm that runs as an infinite loop,
  maximizing the discrete-input MI objective
  $\mathcal{J}_{\bw}(\blam)$ under~\eqref{eq:power_constraint}
  on a millisecond time scale.
\item \textbf{System~2 (slow loop, LLM navigator):}
  An LLM that periodically translates a natural-language \emph{policy}
  into bounded control parameters $(\bw, \Ptot)$, enabling the
  operator to switch operating points without reimplementation.
  It runs on a second-to-multi-second time scale.
\end{itemize}

\noindent
The two systems are \emph{decoupled by time-scale separation}:
System~1 never waits for System~2, and System~2 does not
intervene in individual optimization steps.
System~2 steers System~1 indirectly by reshaping its objective
function (via~$\bw$) and constraint (via~$\Ptot$).

A critical design choice is that the LLM \emph{never directly
manipulates} the allocation variables~$\blam$.
Feasibility ($\sum_i \lambda_i^2 \le \Ptot$) is enforced by
System~1's projection step regardless of the LLM's output.
This \emph{indirect control} structure provides bounded actuation
and graceful degradation under LLM failure.

\subsection{System~1: Projected Gradient Ascent}\label{sec:system1}

System~1 implements the projected gradient ascent outlined in
Section~\ref{sec:grad_ascent} as a continuous loop.
At each iteration, the weighted gradient
$g_i = w_i\,{\partial I_i}/{\partial \lambda_i}$
is obtained via autograd through~\eqref{eq:mi_mc}, and the update rule is given by
\begin{equation}\label{eq:update}
  \lambda_i \leftarrow \lambda_i + \eta\, g_i\,,
\end{equation}
where $\eta > 0$ is the step size.
After each update, each component is first clamped to a positive
lower bound to prevent the \emph{zero-gradient trap}:
$
  \lambda_i \leftarrow \max(\lambda_i,\,\lmin).
$
At $\lambda_i = 0$ the gradient $\partial I_i / \partial \lambda_i$
vanishes, so without clamping a channel that loses all power
could never recover via gradient ascent alone.
By maintaining $\lambda_i \ge \lmin > 0$, every channel retains a
nonzero gradient and can be reactivated when the weights change.
Then the power constraint is enforced by projecting onto
the $\ell_2$-ball of squared radius $\Ptot$:
\begin{equation}\label{eq:projection}
  \|\blam\|_2^2 > \Ptot \;\;\Longrightarrow\;\;
  \blam \leftarrow \blam\,\frac{\sqrt{\Ptot}}{\|\blam\|_2}\,.
\end{equation}
Since the projection is applied \emph{after} clamping, the total power
constraint $\sum_i \lambda_i^2 \le \Ptot$ is satisfied strictly at
every iteration.
Clamping may be partially undone by the subsequent projection, but
for $N\lmin^2 \ll \Ptot$ (e.g., $\lmin = 0.1$, $N = 8$ gives
$N\lmin^2 = 0.08$ versus $\Ptot = 40$) the effect is negligible.

System~1 runs as an infinite loop, repeatedly applying
\eqref{eq:update}--\eqref{eq:projection}.
When the environment $h_i(t)$ or the control parameters $(\bw, \Ptot)$
change, the optimal $\blam^\star$ shifts, and System~1 autonomously
tracks the moving optimum---a form of \emph{homeostasis}.
Crucially, System~1 operates \emph{independently of the LLM}:
if System~2 fails or is disconnected, System~1 continues to optimize
under the last-received parameters, providing
\emph{graceful degradation}.

\subsection{System~2: LLM Navigator}\label{sec:system2}

While System~1 determines \emph{how} to optimize, System~2 determines
\emph{what} to optimize by periodically invoking an LLM.

\subsubsection{Main loop}
System~2 runs an independent loop at a period of
$T_{\mathrm{LLM}}$ seconds (typically 1--3\,s), asynchronously with
System~1.
Each cycle consists of three steps:
(i)~read the current system state from System~1;
(ii)~compose and send a prompt to the LLM;
(iii)~parse the response and apply post-processing
(Section~\ref{sec:guardrails}) before writing the updated
$(\bw, \Ptot)$ back to System~1.
Only one LLM call is in flight at a time; if the previous call has not
returned, the cycle is skipped.

\subsubsection{Input: system prompt and state summary}
Each LLM call sends exactly two messages---a \emph{system prompt}
and a \emph{user message}---with no conversation history
(memoryless design; see below).
The system prompt is structured into four blocks:
(i)~a \emph{system model} block giving the channel equation and the
meaning of each observable;
(ii)~a \emph{control mechanism} block explaining that increasing $w_i$
tends to steer more power toward channel~$i$---this mechanistic
explanation is critical for policy-to-weight translation;
(iii)~an \emph{output specification} block defining the JSON schema; and
(iv)~a \emph{task} block stating the periodic calling convention and
default behavior.
The user message contains the current state summary and policy string.
A representative example (abridged) is:

\smallskip\noindent
{\footnotesize
\begin{verbatim}
Total MI = 12.43 bits, Power = 38.5 / 40.0
 ch |h|^2   MI   P_i   w_i
  1  0.25  1.02  3.20  0.125
  ...
  8  2.25  1.98  8.60  0.125
Policy: Maximize total throughput
\end{verbatim}
}\smallskip

\subsubsection{Output: JSON control action}
The LLM is instructed to respond with a single JSON object:
\texttt{\{``weights'': [$w_1,\dots,w_{N}$],
``P\_total'': $\hat{P}_{\mathrm{total}}$\,(optional),
``reasoning'': ``\dots''\}}.
The \texttt{weights} array specifies the proposed channel priorities,
while \texttt{reasoning} provides a natural-language justification of
the decision (used for explainability; see Section~\ref{sec:guardrails}).

\subsubsection{Indirect control}
The LLM controls $(\bw, \Ptot)$ only; it cannot set $\blam$ directly.
This indirection is the primary safety mechanism:
regardless of the LLM's output, the power
constraint~\eqref{eq:power_constraint} is always satisfied by
System~1's projection~\eqref{eq:projection}.

\subsubsection{Memoryless design}
Each LLM invocation is \emph{stateless}: the message array always
consists of exactly the system prompt and the current user message,
with no prior turns.
The LLM decides based solely on the current snapshot; all temporal
context is embedded in the state summary computed by System~1.
This simplifies context management, bounds token consumption, makes
each call independently recoverable from errors, and keeps the
computational load on the LLM low by avoiding long contexts.
It also enhances explainability: because every decision is a
self-contained function of the current state and policy alone,
the LLM's rationale is easy to inspect.

\subsubsection{Policy persistence and autonomous adaptation}
The \emph{policy} is a declarative statement of the user's intent
(``what to achieve''), not a procedural instruction
(``how to achieve it'').
When the environment changes---e.g., certain channel gains degrade---the
LLM re-evaluates the state against the unchanged policy and adjusts
$(\bw, \Ptot)$ accordingly, \emph{without requiring a new instruction
from the user}.
This policy persistence is the mechanism by which the system exhibits
autonomous adaptation and resilience against abrupt environmental
changes such as sudden channel-gain variations or power-budget shifts.
Table~\ref{tab:policies} lists representative policies used in our
experiments; these are illustrative examples, and any reasonable
natural-language statement of intent can serve as a policy without
modifying the system.

\begin{table}[t]
\centering
\caption{Representative natural-language policies.}
\label{tab:policies}
\footnotesize
\begin{tabular}{@{}c>{\ttfamily\raggedright\arraybackslash}p{5.8cm}@{}}
\toprule
\textbf{ID} & \textrm{\textbf{Policy (natural language)}} \\
\midrule
P1 & Maximize total throughput \\
P2 & Prioritize channels 7 and 8 \\
P3 & Minimize total power while keeping sum rate above 10 bits \\
P4 & Shut down the 3 weakest channels and focus power on the rest \\
\bottomrule
\end{tabular}
\end{table}

\subsection{Multi-Layer Guardrails}\label{sec:guardrails}

Since LLMs are inherently stochastic, integrating one into a control
loop introduces risks from hallucination, parsing errors, and erratic
outputs.
We mitigate these through five layers of guardrails:

\subsubsection{Structural safety via indirect control}
As described in Section~\ref{sec:system2}, the LLM never directly
sets~$\blam$.
Power constraint satisfaction is \emph{mathematically guaranteed} by
System~1's projection~\eqref{eq:projection}, irrespective of the LLM
output.

\subsubsection{Normalization and clamping}
The raw LLM output $\hat{\bw}$ is post-processed:
(i)~negative entries are clipped to zero;
(ii)~the vector is normalized to $\sum_i w_i = 1$;
(iii)~if all entries are zero, a uniform fallback $w_i = 1/N$ is applied.
The power budget $\hat{P}_{\mathrm{total}}$ is clamped to a safe range
$[P_{\min},\, P_{\max}]$.

\subsubsection{EMA smoothing}
To suppress abrupt changes caused by LLM output fluctuations,
updates are smoothed via exponential moving average (EMA):
\begin{align}
  \bw(t{+}1)   &\leftarrow (1-\beta)\,\bw(t)
                  + \beta\,\hat{\bw}(t), \label{eq:ema_w} \\
  \Ptot(t{+}1) &\leftarrow (1-\beta)\,\Ptot(t)
                  + \beta\,\hat{P}_{\mathrm{total}}(t), \label{eq:ema_p}
\end{align}
where $\beta \in (0,1]$ balances damping against hallucination-induced
spikes versus policy tracking speed.

\subsubsection{Fallback and graceful degradation}
On any LLM failure (connection error, parse failure, dimension
mismatch), the current $(\bw, \Ptot)$ is retained and System~1
continues under the last valid parameters---an LLM outage causes
loss of adaptation capability, not system failure.

\subsubsection{Explainability}
The LLM outputs a \texttt{reasoning} field alongside its decisions,
providing a natural-language audit trail for human-in-the-loop
oversight.

Together, these five layers confine the LLM to a bounded semantic
steering role, preserve feasibility by construction, and limit
failures to degraded adaptation rather than unsafe control actions.

\begin{figure*}[!t]
\centering
\includegraphics[width=0.9\textwidth]{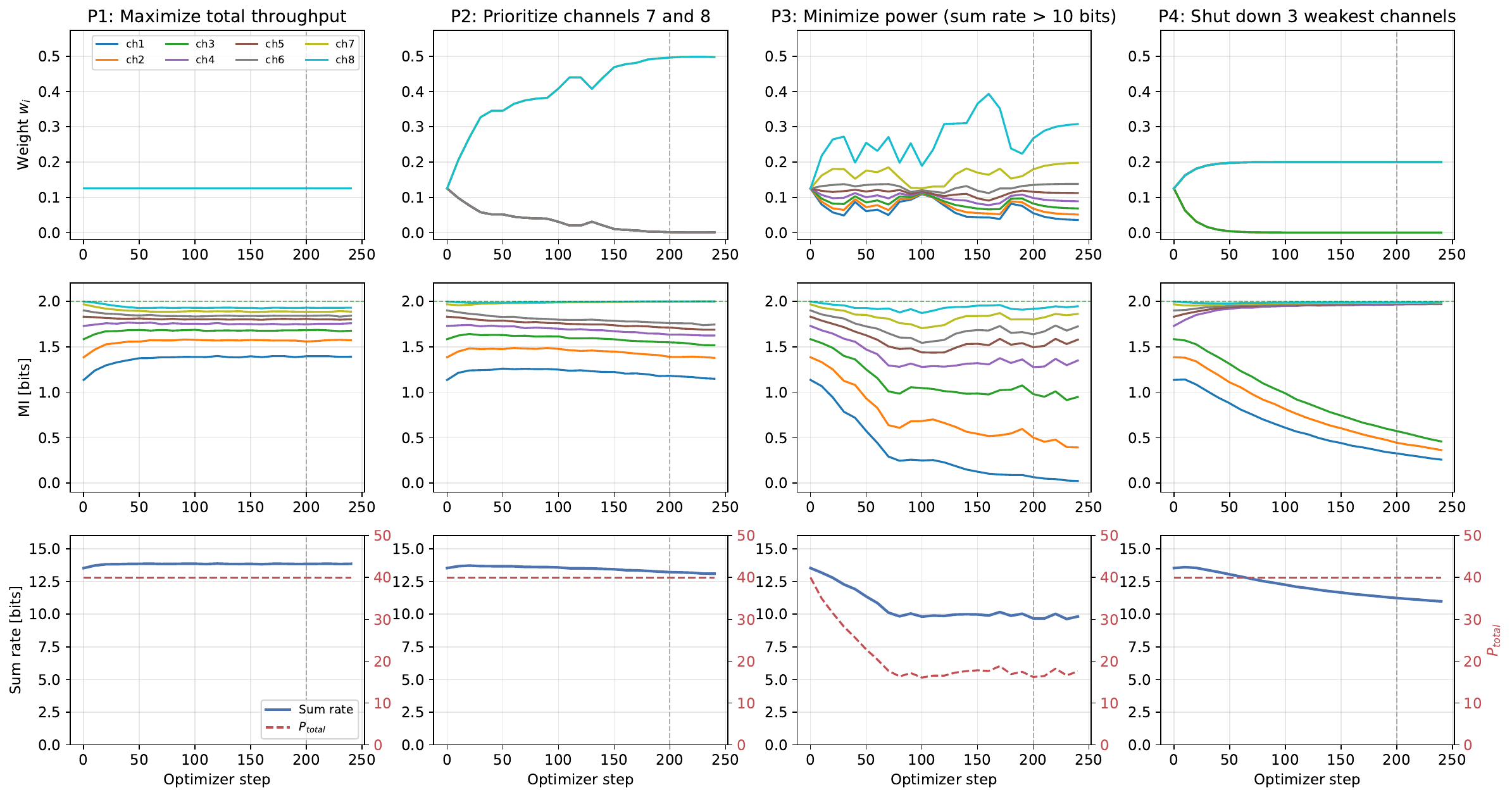}
\caption{Time-series trajectories of LLM-steered power allocation
under four policies (P1--P4 in Table~\ref{tab:policies}).
\textbf{Top:} weight vector $\bw$.
\textbf{Middle:} per-channel MI (bits);
dashed green line $= \log_2 4 = 2$\,bits.
\textbf{Bottom:} sum rate (blue) and $\Ptot$ (red dashed).
Vertical gray dashed line marks end of warmup.}
\label{fig:timeseries}
\end{figure*}

\section{Numerical Experiments}\label{sec:experiments}

\subsection{Setup}\label{sec:exp_setup}

All experiments use $N = 8$ parallel QPSK-AWGN sub-channels with
$\sigma^2 = 1$, default $\Ptot = 40$, and
System~1 parameters $\eta = 1.0$, $\lmin = 0.1$.
The fixed channel gains are
$
  |h_i|^2 = [0.25,\; 0.36,\; 0.49,\; 0.64,\; 0.81,\; 1.0,\; 1.44,\; 2.25].
$
System~2 uses a locally hosted 20\,B-parameter open-weight LLM,
GPT-OSS-20B~\cite{gpt_oss}, served via LM~Studio
(temperature $0.3$, $\mathtt{max\_tokens}=2048$, $\beta = 0.5$)
via the OpenAI-compatible Chat Completions API.
The full system prompt and source code are available in the
accompanying repository (see footnote~1).
For the policy experiment (Section~\ref{sec:exp_policy}),
$\Nmc = 10{,}000$, the LLM is called every 10 optimizer steps,
and each condition runs for 250 steps
(200 warmup $+$ 50 measurement).
Baselines are B0 (System~1 only, $w_i = 1/N$) and
B1 (water-filling~\eqref{eq:waterfilling} evaluated under QPSK MI).
The resilience experiment (Section~\ref{sec:exp_resilience}) uses
$\Nmc = 3{,}000$, a call interval of 20 steps, and 300 total steps.

\subsection{Policy Effectiveness}\label{sec:exp_policy}

Figure~\ref{fig:timeseries} shows the time-series trajectories for
the four policies in Table~\ref{tab:policies}.
Under~\textbf{P1}, the LLM maintained equal weights ($w_i \approx 1/8$),
correctly maximizing sum rate (${\approx}\,13.8$\,bits, matching B0).
Under~\textbf{P2}, the LLM assigned the dominant weight to channel~8
($w_8 \approx 0.5$), while channel~7 received a moderately elevated
weight; both channels reached near-saturation MI
(${\approx}\,2.0$\,bits).
Under~\textbf{P3}, the LLM reduced $\Ptot$ from 40 to
${\approx}\,18$ while concentrating weights on strong channels,
settling the sum rate near the 10-bit target---the only policy that
exercises both control knobs simultaneously.
Under~\textbf{P4}, the three weakest channels were correctly
identified and deactivated ($w_i \approx 0$), yielding near-zero MI
on those channels and ${\approx}\,11$\,bits on the remaining five.

These results show that a \emph{single} LLM with an unchanged system
prompt produces qualitatively distinct, intent-aligned allocation
patterns for each policy---no per-policy reimplementation is needed.
B1 (water-filling, included as a familiar Gaussian-input reference
rather than a discrete-input benchmark) achieved only $13.0$\,bits
versus B0's $13.8$\,bits, confirming the suboptimality discussed in
Section~\ref{sec:waterfilling}.

\subsection{LLM Reasoning Output}\label{sec:exp_reasoning}

Table~\ref{tab:reasoning} shows the reasoning output from the
\emph{first} LLM call under each policy
(Section~\ref{sec:exp_policy}).
Although every call uses the same system prompt and the same
initial system state, the LLM produces a qualitatively different
control strategy for each policy:
P1 correctly identifies that equal weights already maximize
throughput;
P2 concentrates weight on the two designated channels;
P3 is the only policy that simultaneously adjusts
\emph{both} control knobs ($\bw$ and $\Ptot$), proposing to
reduce $\Ptot$ from 40 to~30;
and P4 sets the three weakest channels to zero weight.
These outputs illustrate the explainability benefit of the
\texttt{reasoning} field (Section~\ref{sec:guardrails}):
a human operator can inspect and audit the LLM's rationale and verify that
each policy is interpreted as intended before the system acts.

\begin{table}[t]
\centering
\caption{Initial LLM reasoning for each policy
(Table~\ref{tab:policies}).  All four calls receive the same
system state; only the policy string differs.}
\label{tab:reasoning}
\footnotesize
\begin{tabular}{@{}cp{6.2cm}@{}}
\toprule
 & \textbf{Reasoning (verbatim)} \\
\midrule
P1 & Equal weights maximize total throughput under the current
     power budget. \\
P2 & Prioritize channels 7 and 8 by assigning them higher weights
     while keeping others low to still allow some power
     allocation. \\
P3 & Prioritize high-gain channels to achieve the required
     $>$10 bits with less total power; reduce budget to 30 while
     keeping weights proportional to $|h|^2$. \\
P4 & Zero weights for the three weakest channels (1--3) to shut
     them down; equal positive weight 0.2 for the remaining five
     channels so optimizer reallocates power to them. \\
\bottomrule
\end{tabular}
\end{table}

\subsection{Resilience: Homeostasis under Gain Reversal}
\label{sec:exp_resilience}

We next evaluate the system's ability to maintain \emph{homeostasis}
under an abrupt environmental disturbance.
The policy is ``Equalize MI across all channels,'' intended to
minimize MI disparity among sub-channels.
At step~150 of a 300-step run, all channel gains are reversed
($|h_i|^2_{\text{after}} = |h_{N+1-i}|^2_{\text{before}}$),
so that formerly strong channels become weak and vice versa.
Crucially, the LLM receives \emph{no step number or disturbance
schedule}---it observes only the current state at each call
(every 20~steps, $\beta=0.5$), making its response purely reactive.
As a baseline, we run the same scenario with fixed equal weights
$w_i = 1/N$ (System~1 only, no LLM), under the identical seed and
gain schedule.

\begin{figure}[t]
\centering
\includegraphics[width=\columnwidth]{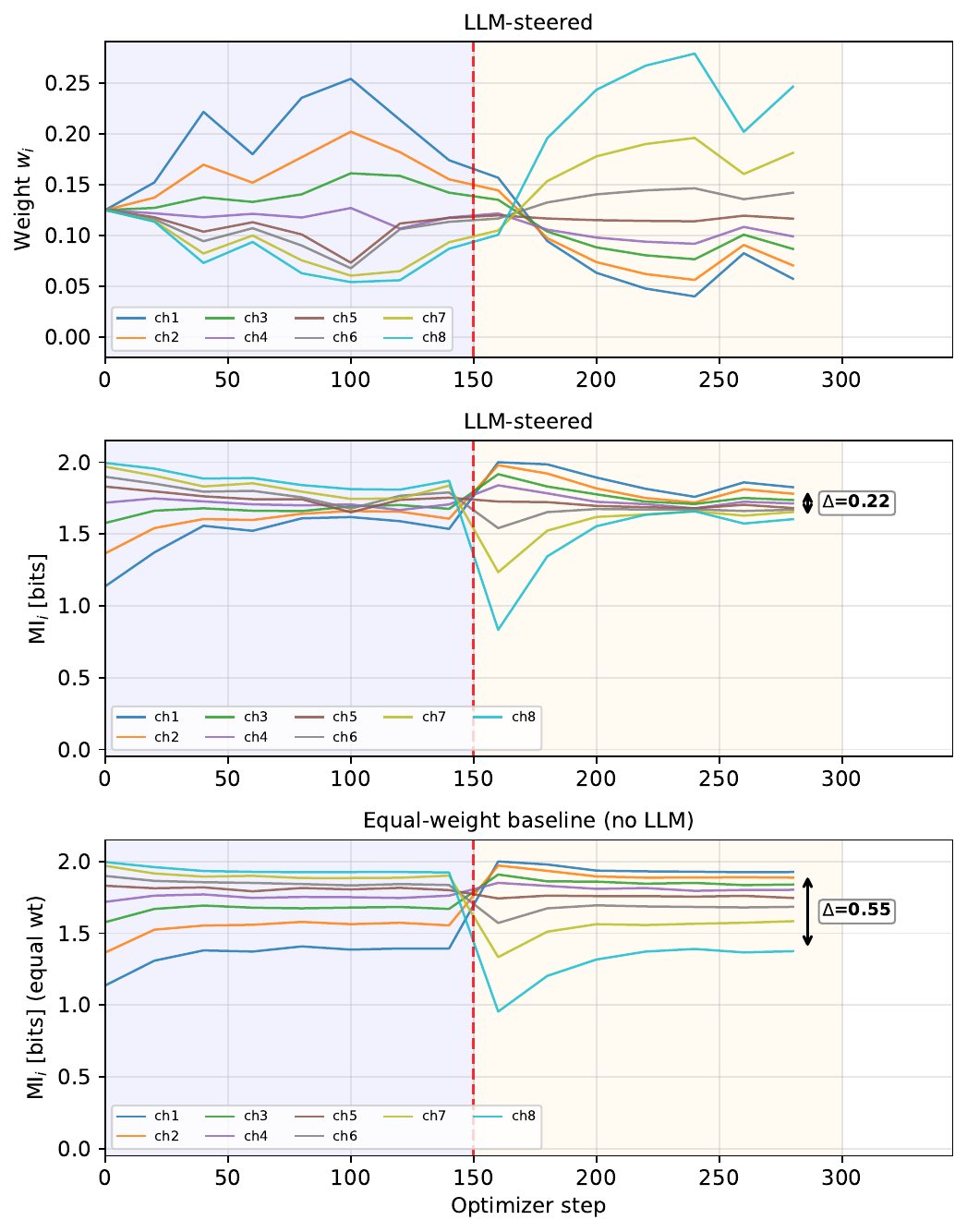}
\caption{Homeostasis experiment.
\textbf{Top:} LLM-steered weights $\bw$;
the LLM boosts weak channels and suppresses strong ones,
flipping after gain reversal (red dashed line, step~150).
\textbf{Middle:} per-channel MI under LLM steering
($\Delta = 0.22$\,bits).
\textbf{Bottom:} equal-weight baseline
($\Delta = 0.55$\,bits).}
\label{fig:resilience}
\end{figure}

Figure~\ref{fig:resilience} shows the results.
Before the reversal, the LLM assigns high weight to the weakest
channel (ch1, $|h_1|^2\!=\!0.25$) and low weight to the strongest
(ch8, $|h_8|^2\!=\!2.25$), progressively equalizing MI.
At step~150 the gains reverse; the LLM detects the change at its
next call (step~160) and flips the weights accordingly---ch8 (now
weakest) receives the highest priority.
We quantify the MI spread at the final step as
$\Delta \triangleq \max_i I_i - \min_i I_i$.
Under LLM steering, $\Delta = 0.22$\,bits, versus
$\Delta = 0.55$\,bits for the equal-weight baseline---a 60\%
reduction.

The baseline (bottom row) isolates System~1's contribution:
the optimizer adapts power allocation to the new gains within
20--40~steps, but under equal weights it maximizes sum rate,
inherently concentrating MI on stronger channels.
LLM steering (System~2) is required to achieve equalization.
This result illustrates the complementary roles of the two
systems and demonstrates resilient, policy-driven homeostasis.

\section{Conclusion}\label{sec:conclusion}

We proposed a dual-process architecture for policy-driven power
allocation in parallel QPSK-AWGN channels, in which an LLM steers
an existing optimizer through a bounded, indirect interface.
Because feasibility is enforced in the optimizer loop, LLM failures
degrade adaptation capability rather than causing unsafe actuation.
Experiments on an 8-channel system showed that a single LLM induces
distinct policy-aligned operating points without controller
reimplementation and reacts to abrupt channel-gain reversal by
reconfiguring its steering signals (60\% MI-spread reduction).
The LLM's \texttt{reasoning} output additionally provides an
operator-facing rationale for each decision, supporting
human-in-the-loop oversight in deployment scenarios.
More broadly, the dual-process paradigm---pairing a fast numerical
solver with an LLM-based semantic reasoner---may provide a useful design pattern for integrating LLM-based
policy interpretation into real-time optimization loops.
Future work includes extension to multi-user MIMO and time-varying
fading channels, as well as investigation of how LLM model size and
capability affect steering quality.

\section*{Acknowledgment}
This work was supported by JST, CRONOS,
Japan Grant Number JPMJCS25N5.

\bibliographystyle{IEEEtran}

\end{document}